# Pantheon 1.0, a manually verified dataset of globally famous biographies

## Authors

Amy Zhao Yu[1], Shahar Ronen[1], Kevin Hu[1], Tiffany Lu[1], César A. Hidalgo[1]

Affiliations

1. Macro Connections, MIT Media Lab

Corresponding author(s): Amy Zhao Yu (ayu@media.mit.edu), César A. Hidalgo (hidalgo@media.mit.edu)

## Abstract

We present the Pantheon 1.0 dataset: a manually verified dataset of individuals that have transcended linguistic, temporal, and geographic boundaries. The Pantheon 1.0 dataset includes the 11,341 biographies present in more than 25 languages in Wikipedia and is enriched with: (i) manually verified demographic information (place and date of birth, gender) (ii) a taxonomy of occupations classifying each biography at three levels of aggregation and (iii) two measures of global popularity including the number of languages in which a biography is present in Wikipedia (*L*), and the Historical Popularity Index (*HPI*) a metric that combines information on *L*, time since birth, and page-views (2008-2013). We compare the Pantheon 1.0 dataset to data from the 2003 book, *Human Accomplishments,* and also to external measures of accomplishment in individual games and sports: Tennis, Swimming, Car Racing, and Chess. In all of these cases we find that measures of popularity (*L* and *HPI*) correlate highly with individual accomplishment, suggesting that measures of global popularity proxy the historical impact of individuals.

## Background & Summary

In this paper, we present a dataset of the biographies of globally famous individuals that can be used to study the production and diffusion of the types of human generated information that is expressed in biographical data by linguistic groups, geographic location, and time period. Biographies allow us to capture people that have either produced a creative oeuvre—such as William Shakespeare and Leonardo da Vinci—or those who have contribute to well known historically events, such as George Washington, a key general in the American Revolution, or Diego Maradona, a key player in Argentina's 1986 World Cup championship.

The Pantheon 1.0 dataset connects occupations, places of births, and dates, helping us create reproducible quantitative measures of the popularity of biographical records that we can use to proxy historical information. This data, therefore, enables researchers to explore the role of polyglots in the global dissemination of human generated information [1], the gender inequality and biases present in online historical information [2], the occupations associated with the producers of historical information, and the breaks produced by communication technologies in the production and dissemination of information by humans [3, 4].

### *Previous efforts*

Past efforts to quantify historical information include Charles Murray's *Human Accomplishments* book, which contributed an inventory of 3,869 significant individuals within the domains of arts and sciences [5]; the digitized text study self-branded by its authors as Culturomics [6]; efforts focused on structuring Wikipedia data [7] and quantifying the impact of individuals across a more diverse set of occupations [8, 9]. Most efforts, however, have looked only at the popularity of individuals in a few languages (predominantly in English) and lack a classification of domains of the contributions that can be used to categorize the areas of historical impact of an individual. This categorization is an essential contribution of our datasets, since without it, it is not possible to study the types of information generated at different time periods and in different geographies.

Table 1 provides a non-exhaustive comparison between various datasets constructed to quantify historical information and the Pantheon 1.0 dataset.

## Methods

### *Data Collection*

Ideally, we would want to quantify historical information by using data that summarizes information produced by people in all languages and that includes all forms of historical information, from biographical data, to the characters created by authors in works of fiction (i.e. Mickey Mouse), and the artifacts and constructions that people create. Since no such dataset exists, we create a simpler dataset focused only on biographical information by using data from Freebase and 277 language editions of Wikipedia. Both Freebase and Wikipedia are open-source, collaborative, multi-lingual knowledge bases freely available online to the general public.

We note that previous efforts have produced structured datasets on biographical records based on Wikipedia [7, 8], but not considering all language editions (using only English), and they have not manually verified time periods and geographies, or introduced a controlled taxonomy of the occupations associated with each biography. While there are certainly considerable limitations to Wikipedia and Freebase, they are currently the largest available domain-independent repositories of collaboratively edited human knowledge, and past research has demonstrated the reliability of these collaborative knowledge bases [10, 11]. We note that we also evaluated Wikidata (http://www.wikidata.org/), the repository for structured data associated with Wikimedia projects, as a possible data source. However, at the time of data collection (2012-2013), this initiative was in its first year of development, and had not yet accumulated a database as robust as what was available within Freebase.

Figure 1 summarizes the main components of the workflow used to create the Pantheon dataset. We derive our dataset of historical biographical information from Freebase's entity knowledge graph (https://developers.google.com/freebase/data) and add metadata from Wikipedia accessible through its API. Freebase organizes information as uniquely identified entities with associated types and properties defined by a structured, but uncontrolled, data ontology. Therefore, to identify globally known biographies, we first determined a list of 2,394,169 individuals through Freebase's database of all entities classified as Persons. Next, we linked individuals to their English Wikipedia page using their unique Wikipedia article id, and from there we obtained information about additional language editions using the Wikipedia API as of May 2013, narrowing the set to the 997,276 individuals that had a presence in Wikipedia. We later supplemented the data with monthly page view data for all language editions from the Wikipedia data dumps for page views for each individual from Jan. 2008 through Dec. 2013.

The Pantheon 1.0 dataset is restricted to the 11,341 biographies with a presence in more than 25 different languages in Wikipedia (L>25). The choice of the L>25 threshold is guided by a combination of criteria, based on the structure of the data and the limits of manual data verification. Figure 2 shows the cumulative distribution of biographies on a semi-log plot, as a function of the number of languages in which each of these biographies has a presence. Most of the 997,276 biographies surveyed have a presence in a few languages, such that the L>25 threshold is a high mark that can help filter the most visible of these biographies. For example, a sampling of the individuals above the L>25 threshold includes globally known individuals such as Charles Darwin, Che Guevara, and Nefertiti. Below the threshold, we find individuals that are locally famous – such as Heather Fargo, who is the former Mayor of Sacramento, California. Also, 95% of individuals passing this threshold have an article in at least 6 of the top 10 spoken languages worldwide (Top 10 spoken languages by number of speakers worldwide: Chinese, English, Hindi, Spanish, Russian, Arabic, Portuguese, Bengali, French, Bahasa – see: http://meta.wikimedia.org/wiki/Top_Ten_Wikipedias), demonstrating that the Pantheon dataset has good coverage of non-Western languages.

### *Taxonomy Design*

Since no globally standardized classification system currently exists to classify biographies to occupations, we introduce a new taxonomy connecting biographies to occupations. Following best practices of taxonomy creation from information science [12], we derive a controlled vocabulary from the raw data and design a classification hierarchy allowing three levels of aggregation. For simplicity we call the most disaggregate level of the taxonomy "occupations," and its aggregation in increasing order of coarseness "industries" and "domains." We note that we use these terms simply to facilitate the communication of the level of aggregation we are referring to. The design and verification process was led by the authors, with support from a multidisciplinary research team with domain expertise in a wide variety of fields, including economics, computer science, physics, design, history, and geography. Figure 3 shows the entire occupation taxonomy, with detail on the all three levels of the classification hierarchy.

To create this taxonomy, we use raw data on individual occupations from Freebase to create a normalized listing of occupations – for example, we map "Entrepreneur", "Business magnate", and "Business development" to the normalized occupation of "Businessperson". We grouped normalized occupations into a second-tier classification (called *industries*), and top-level occupations. We associate individuals within the dataset to a single occupation based on the occupation that best encompasses their primary area of occupation. Thus, we explicitly choose to create a taxonomy—a hierarchical classification that maps each biography to a single category—rather than an ontology—a network that connects biographies to multiple categories—for both technical and historical reasons. The assignment of a biography to multiple categories is troublesome because it requires weighing the multiple classifications that are associated to each biography and defining a threshold for when to stop counting. In some cases the weighing is relatively straightforward. For instance, Shaquille O'Neal should be classified as a basketball player first, and then as an actor, given his 22 actor credits on IMDb. But should we also consider O'Neal a singer (he has released several hip-hop albums), a producer, or a director (he directed a single episode of a little known TV series named Cousin Skeeter)? On a similar note, should we classify Angela Merkel as a physicist and Margaret Thatcher as a chemist (as their respective diplomas indicate), although their historical impact definitely comes from their work in politics? Because of the difficulties involved in defining what categories to consider when assigning an individual to multiple categories, we used a more pragmatic approach and assigned each individual to the category that corresponds to his or her claim to fame: thus, we assign O'Neal to the basketball player category, and assign Merkel and Thatcher to the politician category). By normalizing the data to a controlled vocabulary and using a nested classification system, we provide a consistent mapping for individuals to occupations across time, and enable users of this data to perform analysis at several levels of aggregation while avoiding double counting. Yet we understand that we also introduce the limitation of restricting the contribution of polymaths to one singular domain. The challenge of fairly distributing the historical impact of polymaths will be left for future consideration.

In terms of location assignment, we attribute individuals to a place of birth by country, based on current political boundaries. We use present day political boundaries because of the lack of a historical geocoding API to attribute geographic boundaries using latitude, longitude, and time. Birthplaces were obtained by scraping both Freebase and Wikipedia, and further refined by using fuzzy location matching and geocoding within the Yahoo Placemaker (http://developer.yahoo.com/geo/placemaker/) and Google Maps geocoding (https://developers.google.com/maps/documentation/geocoding/) APIs, and by manual verification. The dataset includes the raw data on individual birthplaces, as well as the cleaned country, which is derived using various APIs that allow us to attribute locations by modern-day country boundaries. To map birthplaces to countries, we normalize the raw data from Freebase indicating the city of birth by latitude and longitude using fuzzy location matching available within the geocoding APIs. Using the coordinates obtained through the APIs, individuals are then mapped to countries based on present-day geographic boundaries using the reverse geocoding API available on geonames.org. For example, individuals born in Moscow during the Soviet Union era are associated with Russia. Using present-day boundaries allows for a consistent basis for matching individuals to countries, and mitigates the technological limitation of the lack of existing historical geocoding APIs for attributing geographic boundaries using latitude, longitude, and time. Historically, birthplace is a fairly suitable way of associating individuals to countries, however, given the increase of human mobility over time [9] and the net migration gains experienced by developed regions [13], future refinement of the dataset may include consideration for improving the attribution of individuals to the geographies he inhabited across his life.

*Visibility Metrics*

We introduce metrics of popularity that help us capture the relative visibility of each biography in our dataset. The fame, or visibility, of historical characters is estimated using two measures. The simpler of the two measures, denoted as $L$, is the number of different Wikipedia language editions that have an article about a historical character. The documentation of an individual in multiple languages is a good first approximation for their global fame because it points to individuals associated with accomplishments or events that have been noted globally. The use of languages as a criterion for inclusion in our dataset helps us differentiate between biographies that are *globally* famous and *locally* famous.

We also introduce the *Historical Popularity Index* (*HPI*), a more nuanced metric for global historical impact that takes into account the following: the individual's age in the dataset ($A$), or the time elapsed since his/her birth, calculated as 2013 minus birthyear; an $L*$ measure that adjusts $L$ by accounting for the concentration of pageviews among different languages (to discount characters with pageviews mostly in a few languages, see Equation 1); the coefficient of variation ($CV$) in pageviews across time (to discount characters that have short periods of popularity); and the number of non-English Wikipedia pageviews ($v^{NE}$) to further reduce any English bias. In addition, to dampen the recency bias of the data, HPI is adjusted for individuals known for less than 70 years. Equation 4 provides the full formula for HPI. There we use log based 4 for the age variable in the aggregation to avoid age becoming the dominant factor in HPI (as it would if we would have used natural log).

For each biography *i*, we define:

$L_i =$ Number of different languages editions of Wikipedia for biography i

$L_i^* =$ Effective number of language editions for biography i

$$L_i^* = \exp(H_i) \tag{1}$$

where $H_i$ is the entropy in terms of Page Views

$$H_i = -\sum_j \left( \frac{v_{ij}}{\sum_i v_{ij}} \ln\left( \frac{v_{ij}}{\sum_i v_{ij}} \right) \right) \tag{2}$$

and $v_{ij} =$ total page views of individual i in language j

$A_i$ = 2013 – Year of Birth

CV = Coefficient of variation in page views

$$CV_i = \frac{\sigma_i}{\mu_i} \tag{3}$$

$\sigma_i$ = standard deviation in pageviews across all languages

$\mu_i$ = average monthly pageviews

$v^{NE}$ = total pageviews in non-English editions of Wikipedia

Using the above, the Historical Popularity Index (HPI) of an individual, i, is defined as:

$$\boldsymbol{HPI} = \begin{cases} \boldsymbol{ln(L) + ln(L^*) + log_4(A) + ln(v^{NE}) - ln(CV)} & \boldsymbol{if\ A \geq 70} \\ \boldsymbol{ln(L) + ln(L^*) + log_4(A) + ln(v^{NE}) - ln(CV) - \frac{70-A}{7}} & \boldsymbol{if\ A < 70} \end{cases} \tag{4}$$

Table 2 shows the ten people with the highest $L$ and $HPI$, respectively, for a few selected periods. An individual is assigned to a period according to his or her date of birth. Here we see that the most notable biographies for each period are associated primarily with well-known historical characters.

## *Biases & Limitations*

As with all large data collection efforts, Pantheon is coupled with limitations and biases, which should be considered carefully when interpreting the dataset. This dataset should be interpreted narrowly, as a view of historical information that emerges from the multilingual expression of historical figures in Wikipedia as of May 2013. The main biases and limitations of the dataset come from:

1. The use of Wikipedia as a data source.
2. The use of place of birth to assign locations.
3. The use of biographies as proxies for historical information.
4. Other technical limitations.

### 1. The use of Wikipedia as a data source

The data is limited by the set of people who contribute to Wikipedia. Wikipedia editors are not considered to be a representative sample of the world population, but a sample of publicly-minded knowledge specialists that are willing and able to dedicate time and effort to contribute to the online documentation of knowledge. Wikipedia editors have an English Bias, a Western Bias, a gender bias towards males, and they tend to be highly educated and technically inclined. They are also more prevalent among developed countries with Internet access. Wikipedia also has a considerable bias in the inclusion of people from different categories. This bias could be the result of the differences in the notability criteria in Wikipedia for biographies from different domains, or from systematic biases within the Wikipedia community [14]. Finally, Wikipedia also has a recency bias, since current events and contemporary individuals typically have greater prominence in the minds of Wikipedia contributors than events from the past [15, 16].

By using data from all Wikipedia language editions we are effectively reducing a bias that would favor information that is locally famous among English speakers. As an example, we note that there is only one American Football Player in the dataset: O.J. Simpson. Certainly, his global notoriety is not purely from his football career, showing that the use of many languages reduces the English bias of the dataset (famous American Football players, such as Peyton Manning, Tom Brady and Joe Montana all have a large presence in the English Wikipedia, but fail to meet the L>25 threshold). In comparison, the dataset contains over 1,000 soccer players – showing that soccer is a sport that is globally popular.

## 2. The use of place of birth to assign locations

Individuals were assigned to geographic locations using their place of birth, based on present-day political boundaries. Country assignments were complemented with geocoding APIs for normalization and manual verification (to correct for errors in API and completeness). Place of birth is one way of assigning a location to an individual that allow us to assign locations in a comprehensive and consistent manner. Yet, there are biases and limitations that need to be considered when using this location assignment method. An important limitation is the inability to account for individuals who became globally known after immigrating to another country. Would Neruda, Picasso or Hemingway be as famous if they had not participated of the Parisian art scene? The place where an individual was born may differ from the place where that individual made his or her more important contributions. In some cases, the contributions are made in a number of different places, and the use of birthplace is unable to capture where the contributions were made. This is particularly true for athletes who migrate to the world's most competitive leagues, or artists that move to the artistic centers of their time. In this dataset, such individuals are not represented since programmatically geo-coding birthplaces is more consistent than registering the place where each individual made his or her more significant contribution, which can only be found through the unstructured data buried in historical narratives.

## 3. Limitations in the use of biographies as proxies for historical information

The use of biographies to proxy historical information allows us to connect information with a linguistic group, geographic location, occupation, and time period. Some biographies involve people that produced an oeuvre directly, like Mozart or Michelangelo, but others reflect important historical events. So, biographies help us capture historical information in a broad sense because they are not limited only to the biographies of those who produced an oeuvre, but because they also include individuals who have inspired documentation by having participated in events that punctuate the history of our species.

The use of biographies as proxies for historical information, however, has important shortcomings. Biographies may fail to capture information on works or events where the participation of groups trumps that of individuals. For example, consider collective enterprises where the accomplishments are the results of teams and not isolated individuals. Examples of accomplishments that are likely to get excluded include the works of music bands or orchestras, or the products produced by a firm, where the accolades collected from accomplishments are connected to a firm, or brand, rather than to an individual. Also, biographies are a proxy of historical information that is biased against works and events that did not result in the widespread fame of their main actors or creators. Moreover, the global popularity of biographies is known to be biased towards the languages that are more central in the global network of translations [1], biasing the estimates of historical information derived from biographical data to the information produced by the speakers of the world's most connected languages.

### 4. Other Technical Limitations

Other biases and limitations include the volatility of Wikipedia and other online resources, which make the results presented here imperfectly reproducible. For example, the Yahoo Placemaker API, which was used for mapping individuals to countries by birthplace, has been deprecated and is no longer publicly available. Also, Freebase will also be retired as of June 2015, and while there are plans to transfer the data to Wikidata, at the time of writing the future availability of Freebase data is undetermined. Finally, the set of included individuals in the Pantheon 1.0 dataset is static and does not reflect events after early 2013 - as such, individuals who only recently rose to global prominence, including Pope Francis and Narendra Modi, are excluded from this dataset.

## Data Records

The Pantheon dataset is publicly available on the Harvard Dataverse Network and can be accessed directly at: https://dataverse.harvard.edu/dataverse/pantheon. The dataset is visualized at http://pantheon.media.mit.edu, a data visualization engine that allows users to dynamically explore the dataset through interactive visualizations.

The data consists of three files – pantheon.tsv, wikilangs.tsv, and pageviews_2008-2013.tsv (Data Citation 1).

The first file, pantheon.tsv, is a flattened tab-limited table, where each row of the table represents a unique biography. Each row contains the following variable fields:

- **name** – name of the historical character (in English)
- **en_curid** – unique identifier for each individual biography, maps to the pageid from Wikipedia. To map to an individual's biography in Wikipedia, use the en_curid field as an input parameter to the following URL: http://en.wikipedia.org/?curid=[en_curid]. We use the English curid as the unique identifier in the Pantheon dataset; we confirmed that all biographies with $L > 25$ as of May 2013 had an entry in the English Wikipedia.
- **countryCode**- ISO 3166-1 alpha2 (based on present-day political boundaries)
- **countryCode3**- ISO 3166-1 alpha3 country code (based on present-day political boundaries)
- **countryName** – commonly accepted name of country
- **continentName** – name of continent
- **birthyear** – birthyear of individual
- **birthcity** – given birthcity of individual
- **occupation** – occupation of the individual
- **industry** – category based on an aggregation of related occupations
- **domain** – category based on an aggregation of related industries
- **gender** – male or female
- **TotalPageViews** – total pageviews across all Wikipedia language editions (January 2008 through December 2013)
- **L_star** – adjusted L (see Appendix for calculation)
- **numlangs** – number of Wikipedia language editions that each biography has a presence in (as of May 2013)
- **StdPageViews** – standard deviation of pageviews across time (January 2008 through December 2013)
- **PageViewsEnglish** – total pageviews in the English Wikipedia (January 2008 through December 2013)
- **PageViewsNonEnglish** – total pageviews in all Wikipedias except English (January 2008 through December 2013)
- **AverageViews** – Average pageviews per language (January 2008 through December 2013)
- **HPI** – Historical Popularity Index (see Equation 4)

The second file, wikilangs.tsv, is a tab-delimited table of all the different Wikipedia language editions that each biography has a presence in. Each row of the table contains the following variables:

- **en_curid** – unique identifier for each individual biography
- **lang** – Wikipedia language code

- **name** – name in the language specified.

To link to the other editions of Wikipedia, use the lang and name parameters in the following URL: http://[lang].wikipedia.org/wiki/[name]

The third file, pageviews_2008-2013.tsv contains the monthly pageview data for each individual, for all the Wikipedia language editions in which they have a presence. Each row of this table includes the following variables:

- **en_curid** – unique identifier for each individual biography
- **lang** – Wikipedia language code
- **name** – English name
- **numlangs** – total number of Wikipedia language editions
- **countryCode3**- ISO 3166-1 alpha3 country code (based on present-day political boundaries)
- **birthyear** – birthyear of individual
- **birthcity** – given birthcity of individual
- **occupation** – occupation of the individual
- **industry** – category based on an aggregation of related occupations
- **domain** – category based on an aggregation of related industries
- **gender** – male or female
- **2008-01 through 2013-12** – total pageviews for the given month (denoted by the column header)

## Technical Validation

### *Comparison with Human Accomplishments dataset*

We compare the Pantheon 1.0 dataset with the Human Accomplishments (HA) dataset, an independent compilation of 3,869 notable people in the arts and sciences from 800BC to AD 1950 [5]. Unlike Pantheon, HA is based on printed encyclopedias and not online sources, but like Pantheon, HA values the presence of a biography in resources in multiple languages. Since HA is restricted to the arts and sciences domains, it does not include politicians like Julius Caesar, religious figures like Jesus, racecar drivers like Ayrton Senna or chess grandmasters like Gary Kasparov. Nevertheless, we find that our data overlaps significantly with the Human Accomplishment data. The Pantheon dataset contains 1,570 (40%) of the entries available in the Human Accomplishment dataset. The HA dataset is more regionally focused than Pantheon, and we find that many of the individuals in the HA dataset are more locally impactful in their respective geographies, and hence, have a presence in fewer languages in Wikipedia. If we lower the threshold of the Pantheon dataset to include biographies existing in 10 or more languages ($L \geq 10$) we would find an overlap of 2,878 biographies, or 74% of the HA dataset.

We also compare the assignment of individuals to their respective occupations within the Pantheon 1.0 dataset to the inventories within the HA dataset. We note that the HA dataset is based on five inventories (art, science, literature, philosophy, and music). From these inventories only the science inventory is disaggregated into additional fields (Chemistry, Biology, Mathematics, Technology, Astronomy, Medicine, Earth Sciences, Physics, and Science—for scientists that do not fit in any of these fields). The smaller number of categories in HA vis-à-vis Pantheon (13 versus 88) means that we cannot create a one-to-one mapping between both categorization systems. Nevertheless, we map each of the HA categories to its most appropriate counterpart in the Pantheon taxonomy. For instance, we map the individuals in the "Medicine" field in the HA dataset to the "Medicine" industry from the Pantheon 1.0 taxonomy, the individuals in the "Chemistry" field from HA to the "Chemist" occupation in the Pantheon taxonomy, and the individuals in the "Literature" inventory to the "Language" industry in the Pantheon 1.0 taxonomy. We find that there is an 84% agreement when comparing the assignment of occupation and industries, and a 95% overall agreement between the datasets when we consider the coarser occupations. Some examples of discrepancies involve Vladimir Lenin, who is categorized as a philosopher in HA but as a politician in Pantheon, and the photographer Ansel Adams, who is categorized as a scientist in HA but is categorized as a Photographer in Pantheon (within the Fine Arts industry and the Arts domain).

Moreover, we find a positive and significant correlation between the measures of historical impact advanced in both of these datasets. HA gives individuals a relative score that measures their impact on their respective domain. Figure 4 shows the correlation between the measures of historical impact in both Pantheon and HA. The historical impact measures in Pantheon correlate with the number of language editions in Wikipedia (L) with an $R^2$=18% (p-value<$2\times10^{-70}$) and with the HPI index with an $R^2$=12% (p-value=$1.6\times10^{-44}$). Note that unlike Pantheon, HA may classify an individual into multiple domains, with a different score for each one: e.g., Galileo Galilei is classified as an astronomer (with a score of 100 that puts him as the most influential astronomer of all time) as well a physicist (ranking fifth with a score of 83).

### *Comparison with External Measures of Accomplishment*

Following an approach similar to the one used in *Human Accomplishment* [5] we also compare measures of individual accomplishments with the Pantheon dataset. Unfortunately, many occupations are not characterized by external metrics of accomplishment that we can associate to individuals, so we restrict our comparison to occupations where measures of individual accomplishment are available – namely, individual sports. The achievements of individual sportsmen and women can be quantitatively expressed through measures such as number of championship titles won or points scored. Here, we focus on Formula-1 drivers, tennis players, swimmers and chess players as independent case studies that we can use to compare our metrics of global visibility.

### *1.* Formula One Racecar Drivers

First we examine the subset of the dataset containing the top 56 Formula-1 drivers, according to the number of languages in which they have a presence in Wikipedia. For each of these drivers we created an additional dataset with the number of Grand Prix Wins, Championships Won, Podiums (number of times in the top 3), Starts, and a dummy variable for Killed in Action (dummy variables are variables used as statistical controls that take values of zero and one). These variables are used to construct a statistical model explaining the multilingual presence of each driver within Wikipedia as well as each driver's Historical Popularity Index. Since Grand Prix Wins, Championships and Podiums are highly collinear—and hence not statistically significant when used together—only Podiums are used in the final model. Since neither L nor HPI can be negative, we link the fame of biographies to the aforementioned variables using an exponential function of the form:

$$y = Ae^{B_1 x_1} e^{B_2 x_2} e^{B_3 x_3}$$

where $x_1$ is the number of podiums, $x_2$ is number of starts, and $x_3$ is an indicator for whether the individual is killed in action.

The first model in Figure 5a explains 54% of the variance in the number of languages in which each Formula-1 driver has a presence in the Wikipedia, showing that for Formula-1 drivers the number of languages in the Wikipedia accurately tracks accomplishments discounted by time. In contrast, when analyzing the same variables with the Historical Popularity Index, we find a model (see Figure 5b) that explains 68% of the variance in the Historical Popularity Index for each Formula-1 driver. The improved fit suggests that the corrections introduced by HPI enhances the L metric and contributes an improved characterization of accomplishment for this sample of individuals.

### *2.* Tennis Players

Next, we conduct a similar analysis for Tennis Players. The Tennis player subset focuses on the top 52 Tennis players according to the number of languages in the Wikipedia and augmented by additional data on each individual - the number of weeks he/she spent as number one in the ATP or WTA, the number of Grand Slam wins, the top rank ever obtained, and the player's gender (Female = 1, Male = 0). We link the fame of biographies for Tennis Players to the aforementioned variables using an exponential function of the form:

$$y = Ae^{B_1x_1}e^{B_2x_2}e^{B_3x_3}e^{B_4x_4}$$

where $x_1$ is the number of weeks at the number one, $x_2$ is the number of Grand Slam wins, $x_3$ is highest rank obtained, and $x_4$ is the variable for gender.

For the number of language presences in Wikipedia (L), we construct a model which explains 34% of the variance in the multilingual presence of each of these individuals in the Wikipedia (Figure 6a). This shows that once again, the number of languages in Wikipedia is a good proxy for individual accomplishments. When we considered HPI, we find an improved model that explains 63% of the variation in HPI (Figure 6b). This further supports the use of HPI as an appropriate proxy for accomplishment, since HPI tracks the degree of achievement for tennis players better than L.

### *3.* Swimmers

We also perform a similar analysis considering Olympic swimmers born after 1950 (n=19). In this case, the model uses the total number of gold medals and gender. We link the fame of swimmers to these variables using an exponential function of the form:

$$y = Ae^{B_1x_1}e^{B_2x_2}$$

where $x_1$ is the number of gold medals, and $x_2$ indicates gender.

In Figure 7a, the model explains 74% of the variance observed in the total number of languages that a swimmer has a presence in Wikipedia, demonstrating that this measure is a good proxy for measuring accomplishment for swimmers. When we perform the analysis for Historical Popularity Index, we find that the model explains 50% of the variance observed in the HPI for swimmers. Figure 7b shows the second model, which shows that HPI is also an appropriate proxy for quantifying accomplishment for swimmers, although in this case, HPI is not superior to L.

### *4.* Chess Players

Finally, we perform another analysis using all of the 30 individuals classified as chess players in the Pantheon dataset. In this case, we use data on each individual's highest ELO ranking attained, gender, total games played, and percentage of wins, losses, and draws. We link the fame of chess players to these variables using an exponential function of the form:

$$y = Ae^{B_1x_1}e^{B_2x_2}e^{B_3x_3}e^{B_4x_4}e^{B_5x_5}e^{B_6x_6}$$

where $x_1$ is the highest ELO ranking attained, $x_2$ indicates gender, $x_3$ is the total games played, $x_4$ is the percentage of wins, $x_5$ is the percentage of losses, and $x_6$ is the percentage of draws.

For the number of language presences in Wikipedia (L), we construct a model that explains 37% of the variance in the multilingual presence of each of these individuals in the Wikipedia (Figure 8a). This further supports using the number of languages in Wikipedia as a proxy for individual accomplishments. Using HPI (Figure 8b), we find a model that explains 53% of the variation in HPI—demonstrating that HPI is an appropriate proxy for accomplishment, with an improved fit for tracking an individual's achievements.

*Discussion*

We introduced a dataset on historical impact that can be used to study spatial and temporal variations in historical information based on biographies that have a presence in more than 25 language editions of Wikipedia. This manually verified dataset allowed us to link historical works and events to places and time. To distinguish between biographies with different levels of visibility we introduce two measures of historical impact: the number of languages in which an individual has a presence in Wikipedia (*L*), and the Historical Popularity Index (*HPI)*. We compared the Pantheon dataset by comparing it against the Human Accomplishments dataset and also compared our measures of global fame and visibility using external data on the accomplishments of Formula One racecar drivers, tennis players, swimmers, and chess players. In all these cases we find a good match between *L*, *HPI*, and the external measures of accomplishment, demonstrating that the measures developed within Pantheon correlate, for these particular occupations, with historical accomplishments. While these case studies are not exhaustive across all occupations, they show that the measures introduced are effective metrics for characterizing historical information across diverse sets of domains, time, and geography. Consider a Formula One racecar driver. Certainly, for a Formula One racer the number of Grand Prix won, or Championships, would be a better metric of accomplishment than the number of languages in Wikipedia. Yet, since Grand Prix won is a metric that applies only for Formula-1 drivers, it cannot be used for basketball players, swimmers, musicians or scientists. While imperfect, the measures based on the online presence of characters in diverse languages are appropriate proxies accomplishment and provide metrics that we can use to compare individuals from different occupations.

## Usage Notes

The Pantheon 1.0 dataset enables quantitative analysis of historical information and already has demonstrated application in testing hypotheses related to the role of polyglots in the global dissemination of information [1], and the extent of online gender inequality and biases [2]. For future analysis, this dataset can motivate a number of potential areas of research investigating the dynamics of historical information across temporal and spatial dimensions. For example, the data can be used in connection with other datasets to empirically assess the connections between economic flourishing and historical information, the dynamics of fame across different domains and geographies, and the dynamics of our species' collective memory.

The data is provided as flat files in tab-separated format, and no additional pre-processing is necessary for users to import the files into a scientific computing environment. A wide variety of software tools for data visualization and numerical analysis can be used to explore the dataset, including MATLAB, R, the SciPy stack, d3, d3plus, etc. In addition, the data includes a number of fields that can be linked to external datasets, such as standardized country and language codes, and unique individual ids from Wikipedia. We emphasize that future results should be interpreted within the narrow context of the dataset documented, and that analyses of the dataset should include consideration of its bias and limitations.

## Acknowledgements

We are grateful to Defne Gurel, Francine Loza, Daniel Smilkov, and Deepak Jagdish for their contributions and feedback during the data verification and cleaning process. We also want to thank Ethan Zuckerman and Alex Lex for their feedback and comments over the course of this project. This research is supported by funding from the MIT Media Lab Consortia and the Metaknowledge network at the University of Chicago.

## Author Contributions

AZY collected, verified, and compared the data, and wrote the manuscript.
SR contributed to data collection, cleaning, and comparisons, and edited the manuscript.
KH contributed to data cleaning, and supplemented the dataset with pageviews data.
TL contributed to the initial scripts for scraping Freebase and Wikipedia.
CAH conceived the study, verified & compared the data, and wrote the manuscript.

## Data Citation

# Figures

Figure 1: Pantheon Data Workflow

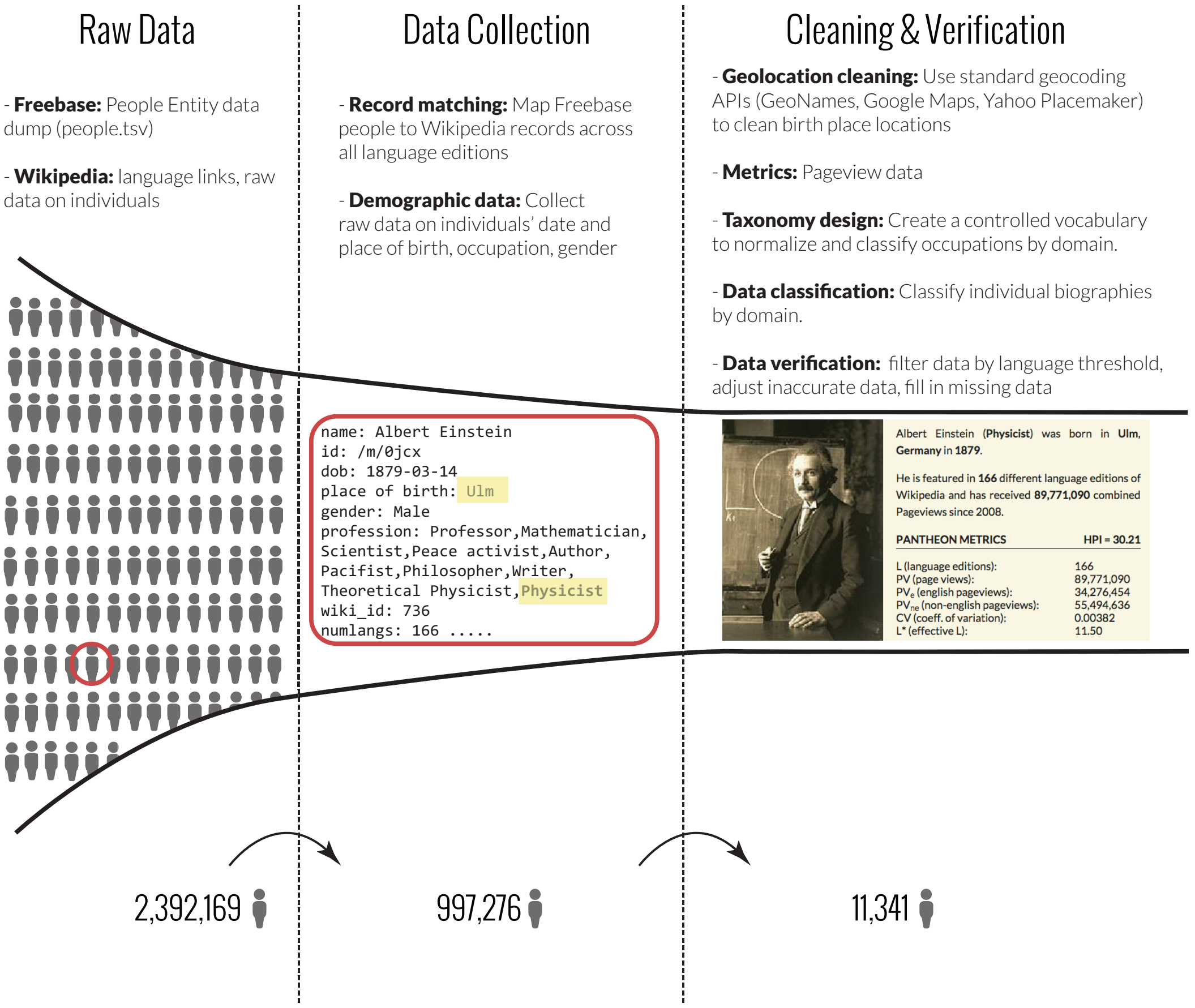

*Flow diagram detailing the data collection process for the Pantheon 1.0 (n=11,341). Inset image from pantheon.media.mit.edu.*

Figure 2: Cumulative Number of Individuals with at least N Wikipedia Language Editions

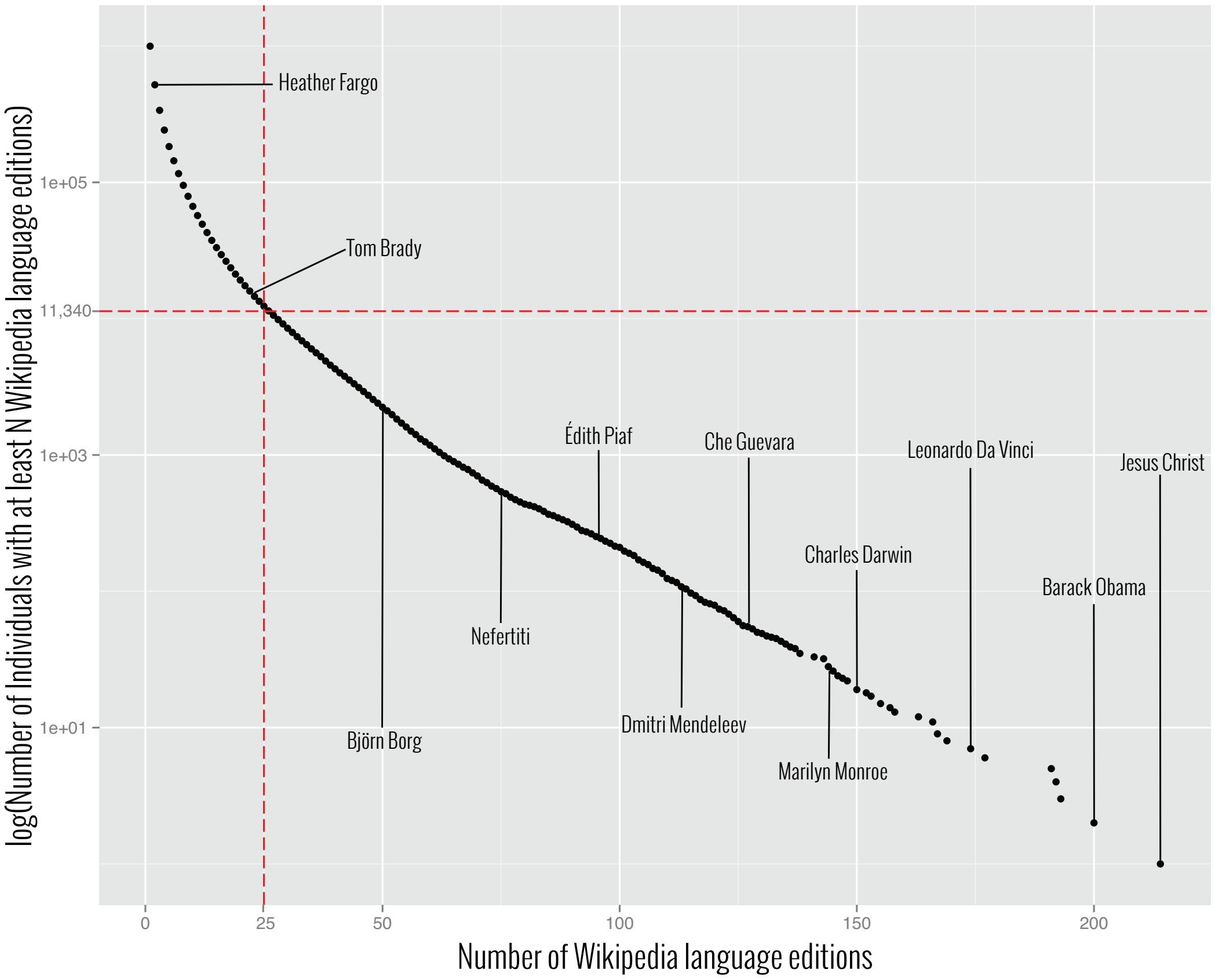

*Cumulative distribution of biographies on a semi-log plot, as a function of the number of languages in which each of these biographies has a presence. Individual images from Wikimedia Commons.*

Figure 3: Domain Taxonomy

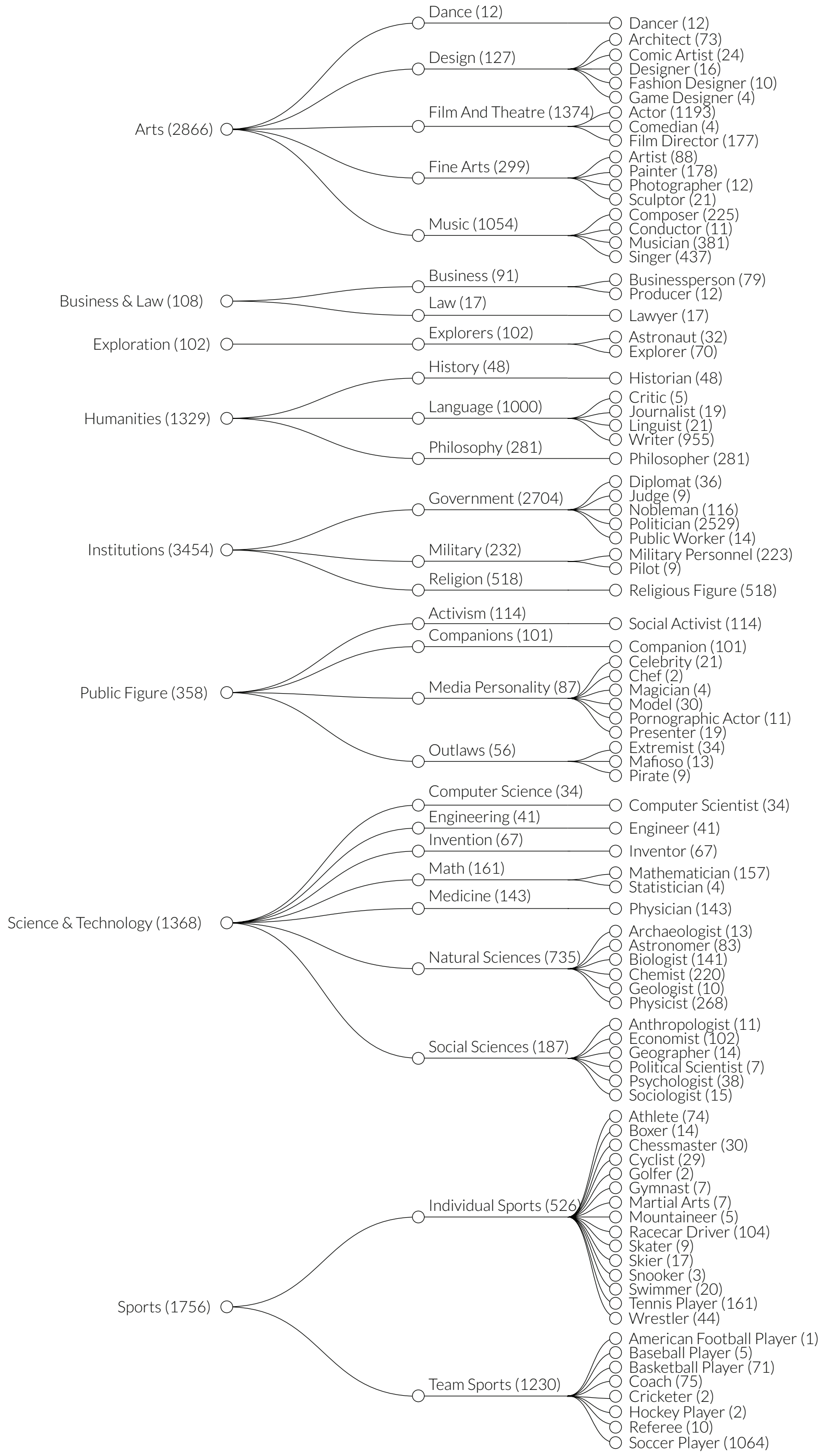

*From left to right: domain (i.e. Sports), industry (i.e. Team Sports) and occupation (i.e. Soccer Player)*

Figure 4: Comparison Analysis of Human Accomplishments and Pantheon 1.0

| name | L | MurrayIndex |
|---|---|---|
| Paul Valéry | 38 | 19.00 |
| Arnold Sommerfeld | 36 | 20.36 |
| Soseki Natsume | 37 | 60.08 |
| Frederick Hopkins | 34 | 20.27 |
| Giuseppe Peano | 40 | 23.78 |
| Alexis Carrel | 43 | 35.66 |
| Hugo de Vries | 31 | 43.66 |
| Giovanni Schiaparelli | 44 | 21.08 |
| James McNeill Whistler | 35 | 19.93 |
| Richard Dedekind | 40 | 23.13 |

| name | L | MurrayIndex |
|---|---|---|
| Pablo Picasso | 143 | 77.32 |
| Marie Curie | 130 | 40.78 |
| Max Planck | 103 | 33.23 |
| Sigmund Freud | 134 | 34.46 |
| Leo Tolstoy | 123 | 40.53 |
| Fyodor Dostoyevsky | 113 | 40.20 |
| Richard Wagner | 116 | 79.55 |
| Dante Alighieri | 137 | 62.50 |
| Archimedes | 129 | 51.14 |
| Thomas Aquinas | 103 | 39.04 |

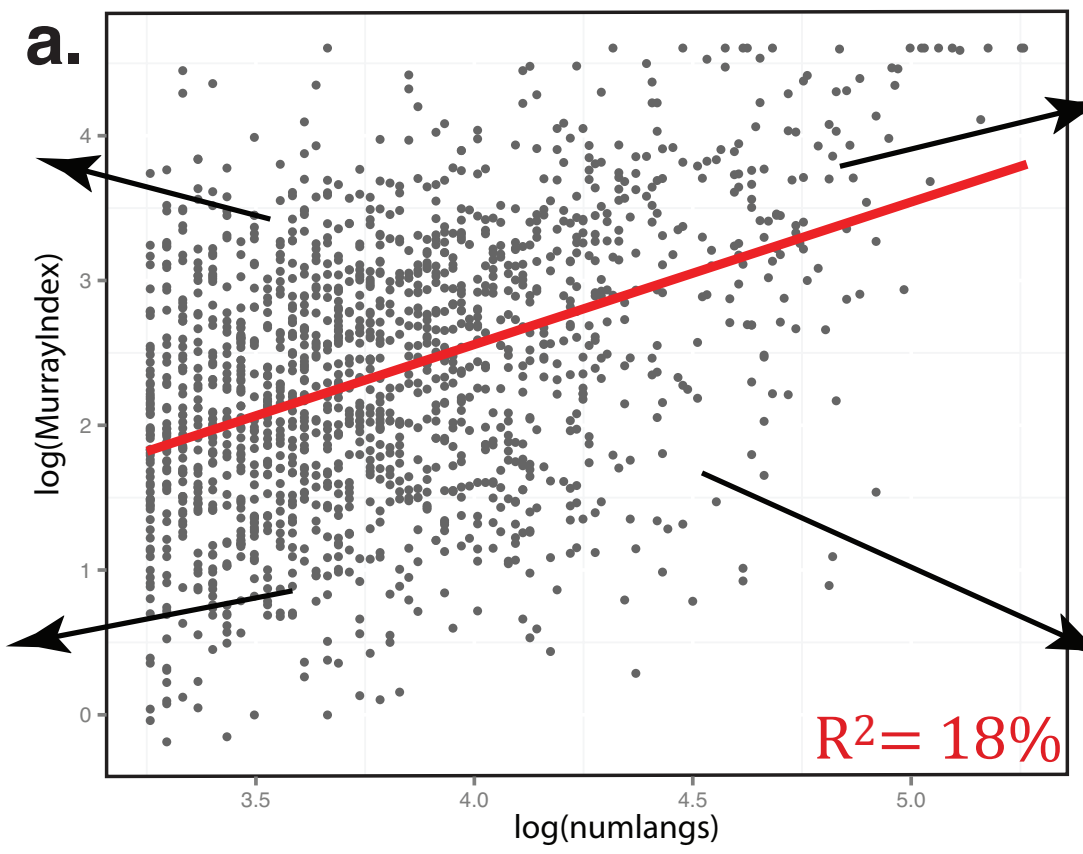

| name | L | MurrayIndex |
|---|---|---|
| Alfred Jarry | 32 | 3.35 |
| Henri Barbusse | 35 | 1.97 |
| Heinrich Mann | 39 | 4.25 |
| Jules Bordet | 39 | 1.66 |
| Fritz Pregl | 42 | 2.94 |
| Paul Dukas | 34 | 3.96 |
| Alexander Glazunov | 36 | 3.51 |
| Carl Nielsen | 35 | 3.20 |
| Max Bruch | 35 | 2.02 |
| Percival Lowell | 32 | 4.39 |

| name | L | MurrayIndex |
|---|---|---|
| Simone de Beauvoir | 85 | 3.61 |
| Pablo Neruda | 95 | 4.35 |
| George Orwell | 101 | 2.52 |
| Arthur Conan Doyle | 90 | 2.19 |
| Victor Hugo | 123 | 2.44 |
| Blaise Pascal | 103 | 6.03 |
| Giordano Bruno | 83 | 3.82 |
| Jorge Luis Borges | 101 | 2.75 |
| Thales | 84 | 6.07 |
| Benjamin Franklin | 88 | 3.73 |

| name | HPI | MurrayIndex |
|---|---|---|
| William Shockley | 23.85 | 16.35 |
| Jacques Monod | 23.46 | 15.05 |
| Edward Lawrie Tatum | 22.70 | 16.69 |
| Alfred Hershey | 22.62 | 13.57 |
| George Gabriel Stokes | 24.19 | 20.94 |
| Herman Melville | 23.93 | 14.50 |
| Jean-Martin Charcot | 24.18 | 24.95 |
| Francis Galton | 24.41 | 26.01 |
| Thomas Carlyle | 24.17 | 13.89 |
| John Herschel | 24.45 | 26.81 |

| name | HPI | MurrayIndex |
|---|---|---|
| Henri Matisse | 27.39 | 38.30 |
| Marie Curie | 27.60 | 40.78 |
| Max Planck | 27.32 | 33.23 |
| Sigmund Freud | 29.63 | 34.46 |
| Vincent van Gogh | 29.74 | 39.76 |
| Claude Monet | 28.50 | 40.70 |
| Édouard Manet | 27.26 | 33.52 |
| Leo Tolstoy | 28.21 | 40.53 |
| Leonardo Fibonacci | 27.83 | 33.71 |
| Laozi | 29.15 | 68.63 |

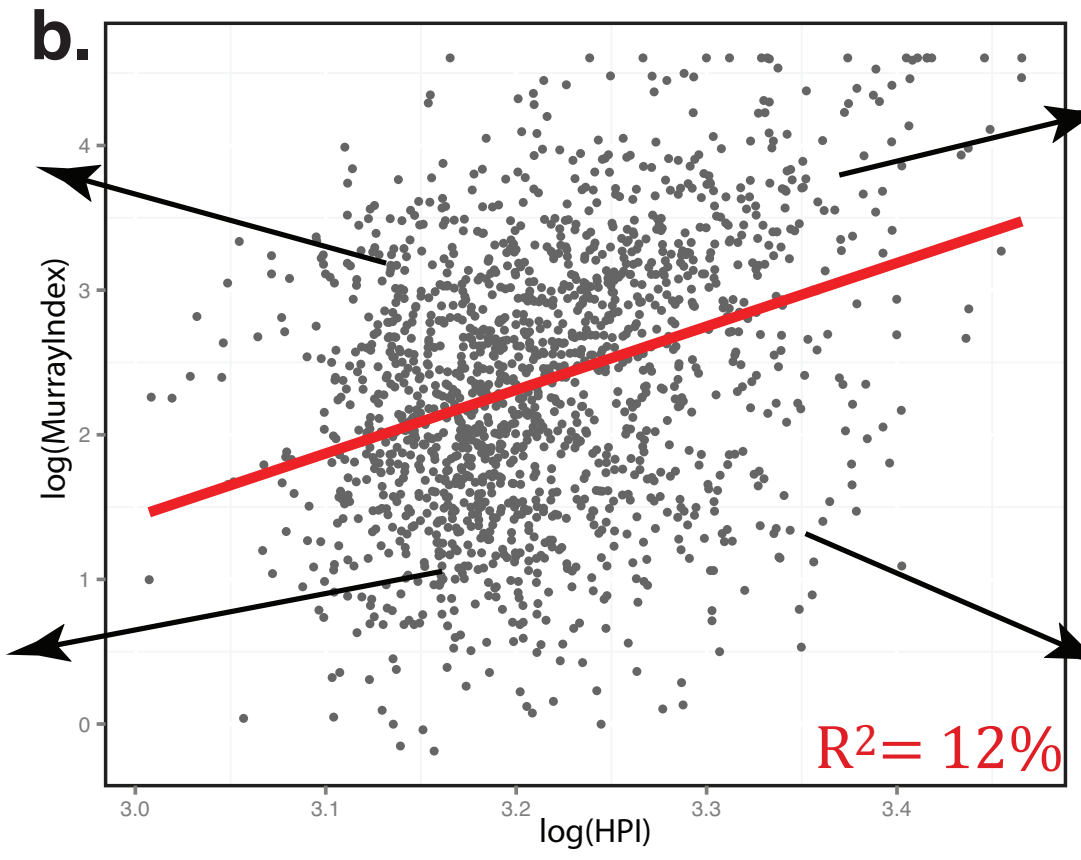

| name | HPI | MurrayIndex |
|---|---|---|
| William Saroyan | 23.08 | 3.54 |
| Albert Sabin | 22.76 | 4.37 |
| Witold Gombrowicz | 23.13 | 1.99 |
| Erskine Caldwell | 22.34 | 4.01 |
| Evelyn Waugh | 22.29 | 2.90 |
| Giulio Natta | 23.33 | 4.08 |
| Rafael Alberti | 23.15 | 2.74 |
| Anna Seghers | 23.09 | 2.46 |
| Georg von Békésy | 22.85 | 3.43 |
| E. E. Cummings | 22.57 | 1.88 |

| name | HPI | MurrayIndex |
|---|---|---|
| Simone de Beauvoir | 27.08 | 3.61 |
| Pablo Neruda | 27.23 | 4.35 |
| George Orwell | 27.66 | 2.52 |
| Benjamin Franklin | 28.09 | 3.73 |
| Seneca the Younger | 28.82 | 4.06 |
| Anaximenes of Miletus | 27.37 | 3.53 |
| Xenophanes | 27.21 | 2.89 |
| Anaximander | 28.12 | 3.87 |
| John Calvin | 28.18 | 3.15 |
| Thomas More | 28.16 | 3.17 |

*(a) Comparison of L (number of language presences in Wikipedia of an individual) with index scores from the Human Accomplishments (HA) dataset, including sample details on individuals matched between the datasets (n= 1,570).*

*(b) Comparison of HPI (Historical Popularity Index) with the index scores from HA, including sample details on individuals matched between the datasets (n= 1,570).*

Figure 5: Analysis with Formula 1 Drivers

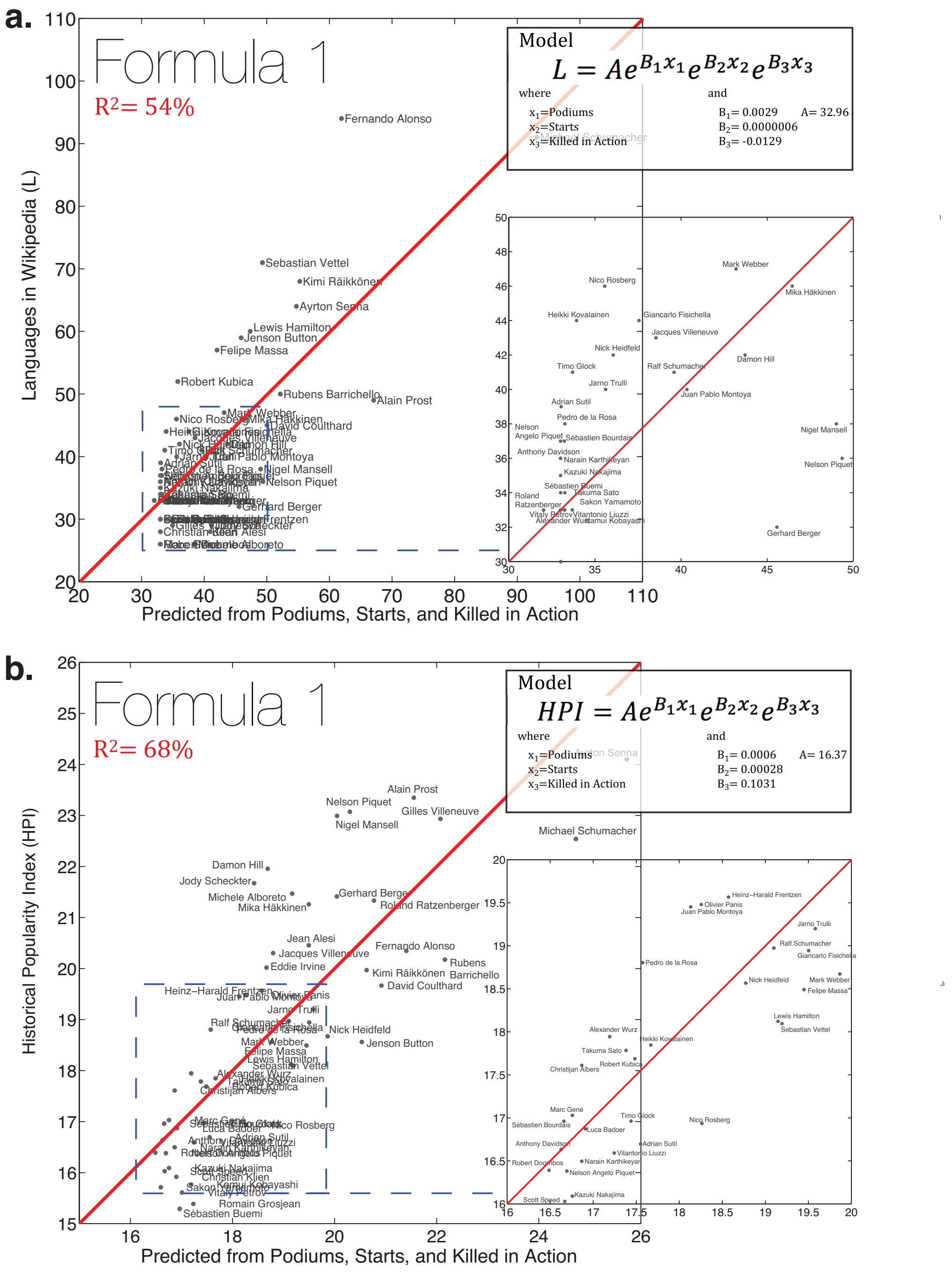

*(a) Analysis of L (number of language presences in Wikipedia of an individual) using Podiums, Starts, and Killed in Action data on Formula 1 drivers (n=56).*

*(b) Analysis of HPI (Historical Popularity Index) using Podiums, Starts, and Killed in Action data on Formula 1 drivers (n=56).*

Figure 6: Analysis with Tennis Players

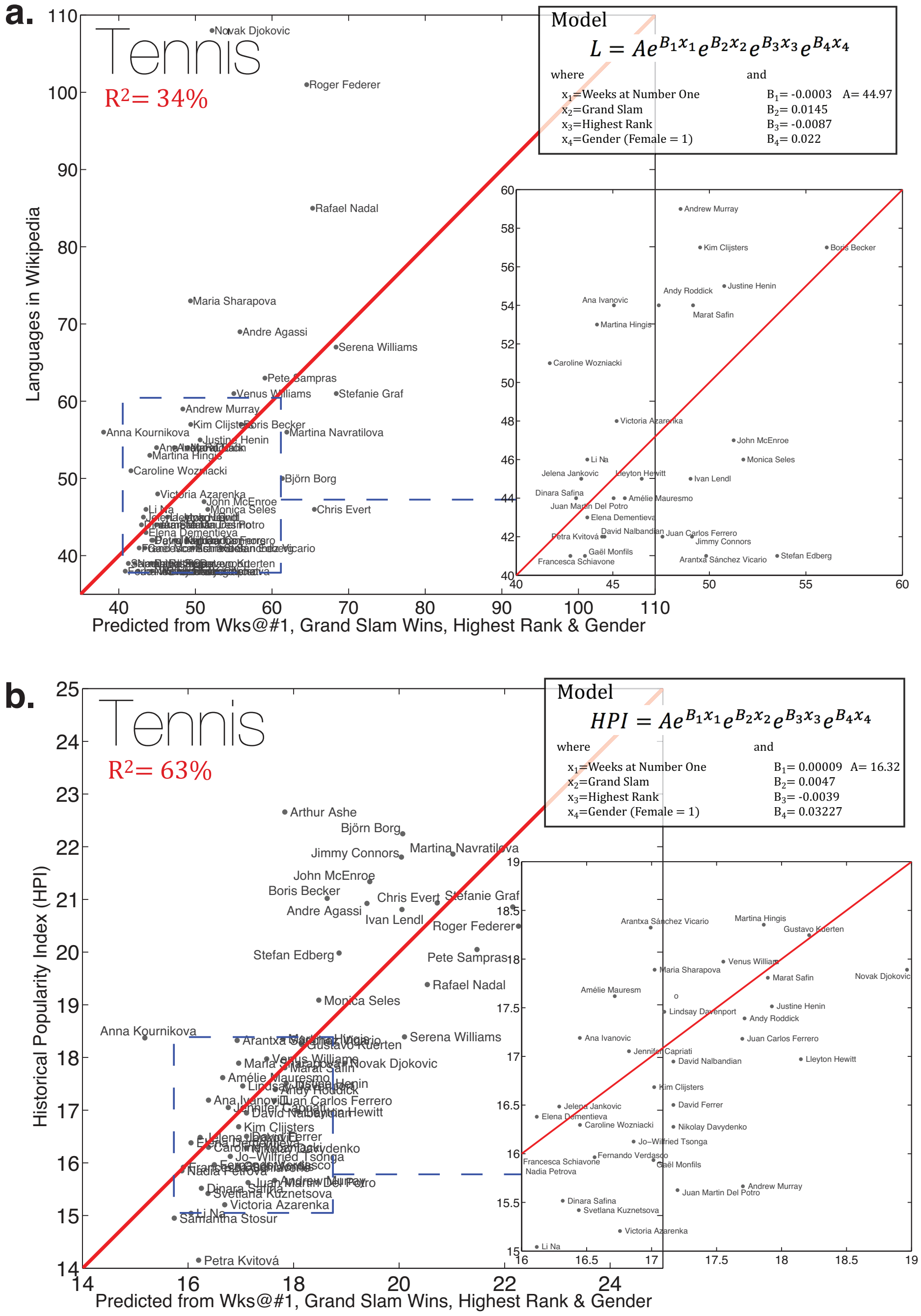

*(a) Analysis of L (number of language presences in Wikipedia of an individual) using number of weeks at number one in the ATP or WTA, the number of Grand Slam wins, the top rank ever obtained, and gender (Female = 1, Male = 0) data on tennis players (n=52).*

*(b) Analysis of HPI (Historical Popularity Index) using number of weeks at number one in the ATP or WTA, the number of Grand Slam wins, the top rank ever obtained, and gender (Female = 1, Male = 0) data on tennis players (n=52).*

Figure 7: Analysis with Swimmers

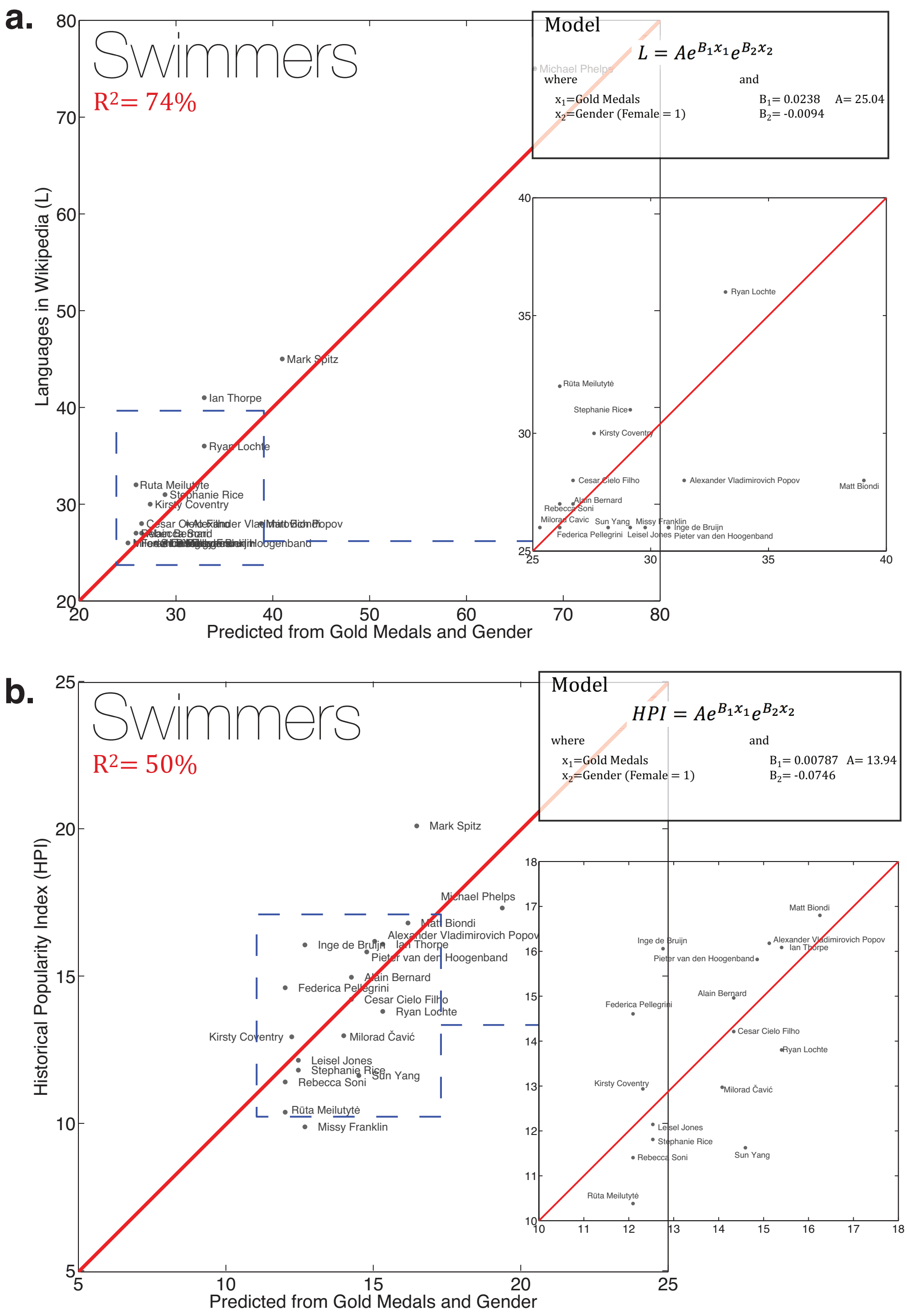

*(a) Analysis of L (number of language presences in Wikipedia of an individual) using the total number of gold medals won and gender data on swimmers born after 1950 (n=19).*

*(b) Analysis of HPI (Historical Popularity Index) using the total number of gold medals won and gender data on swimmers born after 1950 (n=19).*

Figure 8: Analysis with Chess Players

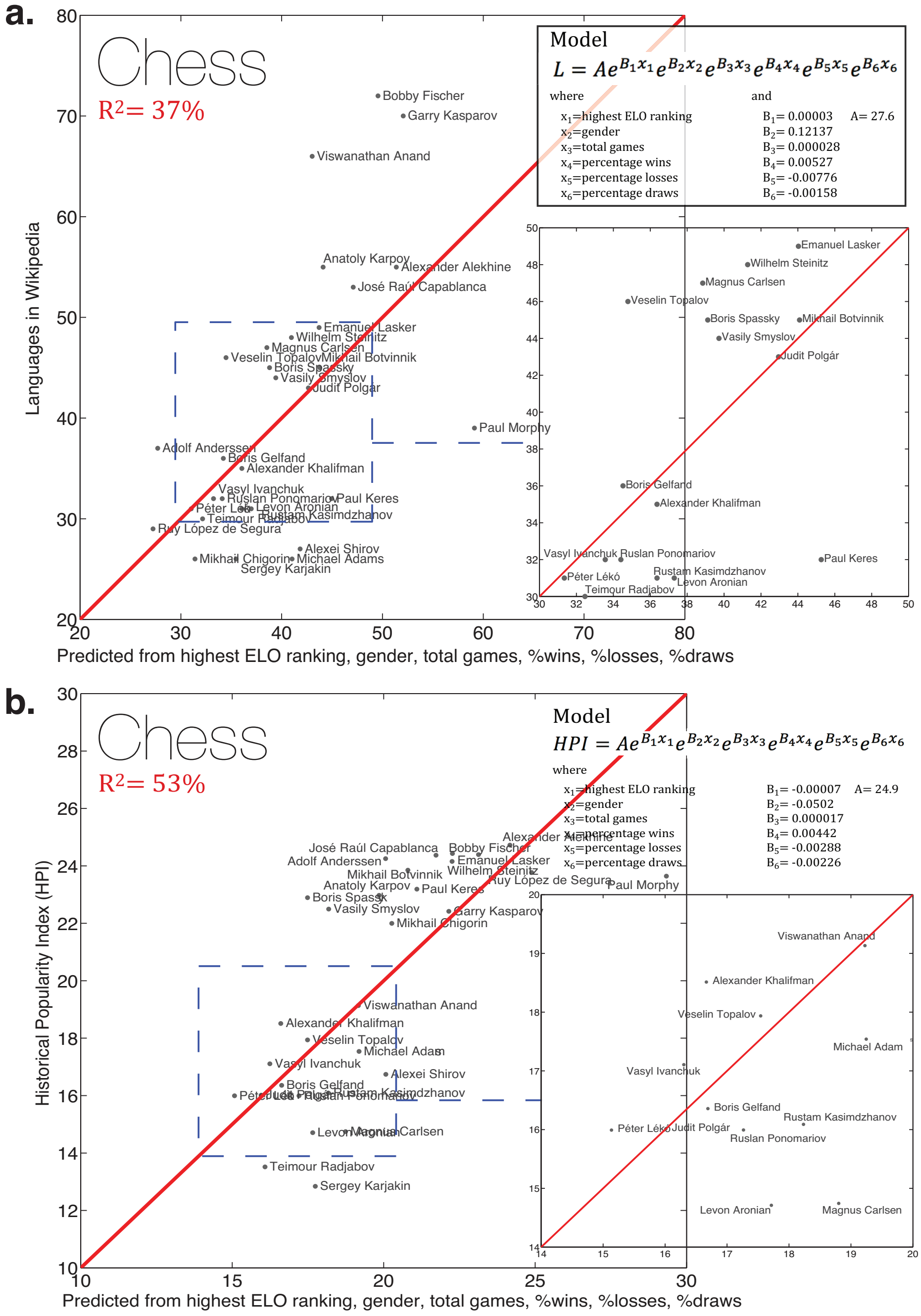

*(a) Analysis of L (number of language presences in Wikipedia of an individual) using highest ELO ranking attained, gender, total games played, and percentage of wins, losses, and draws of chess players (n=30).*

*(b) Analysis of HPI (Historical Popularity Index) using highest ELO ranking attained, gender, total games played, and percentage of wins, losses, and draws of chess players (n=30).*

## Tables

Table 1: Comparison Chart of Quantitative Datasets for Studying Historical Information

| Dataset | Metric(s) | Size | Domains | Includes domain classification | Time | # of Countries | # of Languages |
|---|---|---|---|---|---|---|---|
| UNESCO Cultural Statistics [17] | economic productivity, cultural participation | 52 countries – feature films, 20 countries – employment survey | Arts, Design, Media | Yes (7 core domain headings) | 2000-present | >50 | 2471 |
| World Values Survey (http://www.worldvaluessurvey.org/) | Values | 85,000 respondents | Religion, Institutions | N/A | 1981-present | 87 | 20 |
| Culturomics [6] | *n*-grams | 5,195,769 books | Arts, Politics, Sciences | N/A | 1800-2000 | not reported | 7 |
| Human Accomplishment [5] | Individuals and events | 3,869 significant individuals, 1,560 significant events | Arts & Sciences | Yes (9 within Sciences, 4 within Arts) | 800BC-1950 | 30* countries & regions | 6* languages |
| Who's Bigger [8] | Entities (Individuals, events, places, things) | 843,790 individuals | All domains known in English Wikipedia | Yes (based on English Wikipedia categories) | 2686 BC - 2010 | >200* | 1 (English) |
| Popescu & Grefenstette [7] | Individuals, places | 500,896 names | All domains known in English Wikipedia | Yes (based on English Wikipedia categories) | <600 BC - 2009 | >200* | 7 |
| Networked Framework of Cultural History [9] | Individuals & Places | Multiple datasets (FB, AKL, ULAN, WCEN, Google Ngrams) | All domains | Yes – # categories per dataset: AKL: 5, ULAN: 7, FB: 6 | 1-2012 | >200* | 2 – (mainly English, German) |
| Pantheon 1.0 | Individuals, Places. | 11,341 notable individuals | All domains | Yes (88 occupations, 27 industries, 8 domains) | 4000BC – 2010 | 194 | 277 |

**estimated (exact numbers not reported)*

Table 2: Top 10 Biographies for each Time Period by Number of Language Editions (L) and Historical Popularity Index (HPI)

| Time Period | Top 10 By L | | Top 10 by HPI | |
|---|---|---|---|---|
| | Name | L | Name | HPI |
| Before 500 | Jesus Christ | 214 | Aristotle | 31.99 |
| | Confucius | 192 | Plato | 31.99 |
| | Aristotle | 152 | Jesus Christ | 31.90 |
| | Qin Shi Huang | 144 | Socrates | 31.65 |
| | Plato | 142 | Alexander the Great | 31.58 |
| | Homer | 141 | Confucius | 31.37 |
| | Alexander the Great | 138 | Julius Caesar | 31.12 |
| | Socrates | 137 | Homer | 31.11 |
| | Archimedes | 129 | Pythagoras | 31.07 |
| | Julius Caesar | 128 | Archimedes | 30.99 |
| 500-1199 | Muhammad | 150 | Muhammad | 30.65 |
| | Genghis Khan | 121 | Charlemagne | 30.48 |
| | Charlemagne | 116 | Genghis Khan | 29.74 |
| | Saladin | 104 | Saladin | 29.14 |
| | Avicenna | 102 | Avicenna | 28.89 |
| | Li Bai | 102 | Ali | 28.09 |
| | Muhammad ibn Musa al-Khwarizmi | 100 | Li Bai | 28.07 |
| | Du Fu | 97 | Francis of Assisi | 28.00 |
| | Umar | 92 | Du Fu | 27.91 |
| | Omar Khayyám | 86 | Averroes | 27.83 |
| 1200-1499 | Leonardo da Vinci | 174 | Leonardo da Vinci | 31.46 |
| | Michelangelo | 158 | Michelangelo | 30.44 |
| | Christopher Columbus | 153 | Christopher Columbus | 30.18 |
| | Dante Alighieri | 137 | Dante Alighieri | 30.15 |
| | Nicolaus Copernicus | 128 | Martin Luther | 30.03 |
| | Marco Polo | 127 | Marco Polo | 29.77 |
| | Martin Luther | 125 | Jeanne d'Arc | 29.56 |
| | Ferdinand Magellan | 120 | Thomas Aquinas | 29.44 |
| | Vasco da Gama | 119 | Niccolò Machiavelli | 29.27 |
| | Albrecht Dürer | 112 | Raphael | 29.21 |
| 1500-1749 | Isaac Newton | 191 | William Shakespeare | 30.44 |
| | William Shakespeare | 163 | Isaac Newton | 30.29 |
| | Galileo Galilei | 146 | Johann Sebastian Bach | 30.17 |
| | Johann Sebastian Bach | 144 | Galileo Galilei | 29.96 |
| | George Washington | 134 | Immanuel Kant | 29.69 |
| | Johann Wolfgang von Goethe | 132 | René Descartes | 29.45 |
| | Miguel de Cervantes | 128 | Johann Wolfgang von Goethe | 29.34 |
| | Immanuel Kant | 125 | Voltaire | 29.27 |
| | Rembrandt | 125 | Blaise Pascal | 29.25 |
| | Carl Linnaeus | 123 | Jean-Jacques Rousseau | 29.16 |
| 1750-1849 | Wolfgang Amadeus Mozart | 177 | Wolfgang Amadeus Mozart | 30.51 |
| | Ludwig van Beethoven | 153 | Napoleon Bonaparte | 30.33 |
| | Karl Marx | 148 | Ludwig van Beethoven | 30.11 |
| | Charles Darwin | 148 | Karl Marx | 29.84 |
| | Napoleon Bonaparte | 145 | Charles Darwin | 29.20 |
| | Abraham Lincoln | 131 | Friedrich Nietzsche | 28.67 |
| | Thomas Edison | 126 | Victor Hugo | 28.66 |
| | Victor Hugo | 123 | Richard Wagner | 28.57 |
| | Leo Tolstoy | 123 | Claude Monet | 28.50 |
| | Friedrich Nietzsche | 117 | Frédéric Chopin | 28.47 |
| 1850-1899 | Adolf Hitler | 169 | Adolf Hitler | 30.58 |
| | Huang Xian Fan | 167 | Albert Einstein | 30.21 |
| | Albert Einstein | 166 | Vincent van Gogh | 29.74 |
| | Mustafa Kemal Atatürk | 166 | Sigmund Freud | 29.63 |
| | Vincent van Gogh | 155 | Pablo Picasso | 29.60 |
| | Charlie Chaplin | 145 | Mahatma Gandhi | 29.14 |
| | Pablo Picasso | 143 | Joseph Stalin | 28.93 |
| | Mahatma Gandhi | 138 | Vladimir Lenin | 28.92 |
| | Vladimir Lenin | 137 | Benito Mussolini | 28.55 |
| | Joseph Stalin | 134 | Oscar Wilde | 28.53 |
| 1900-1950 | Hebe Camargo | 157 | Che Guevara | 29.16 |
| | Marilyn Monroe | 143 | Martin Luther King, Jr. | 28.69 |
| | George Bush | 143 | Elvis Presley | 28.62 |
| | Lech Wałęsa | 135 | Salvador Dalí | 28.59 |
| | Nelson Mandela | 133 | Marilyn Monroe | 28.35 |
| | Che Guevara | 129 | Walt Disney | 28.17 |
| | Elizabeth II of the United Kingdom | 127 | Jean-Paul Sartre | 28.15 |
| | Pope Benedict XVI | 123 | Jimi Hendrix | 27.93 |
| | Salvador Dalí | 122 | Andy Warhol | 27.92 |
| | Neil Armstrong | 121 | Mother Teresa | 27.86 |